\documentclass[twocolumn]{aastex631}

\DeclareRobustCommand{\apc}[1]{{#1}}

\DeclareRobustCommand{\replyca}[1]{\textcolor{black}{#1}}
\DeclareRobustCommand{\replyjn}[1]{\textcolor{black}{#1}}

\DeclareRobustCommand{\msun}{\mathrm{M_\odot}}
\DeclareRobustCommand{\gse}{GS/E}

\begin{document}

\title{Progenitor diversity in the accreted stellar halos of Milky Way–like galaxies}

\author[0009-0000-3146-7154]{Sy-Yun Pu}
\affiliation{Institute of Astronomy and Department of Physics, National Tsing Hua University, Hsinchu 30013, Taiwan}
\affiliation{Institute of Astronomy and Astrophysics, Academia Sinica, No. 1, Section 4, Roosevelt Road, Taipei 10617, Taiwan}

\author[0000-0001-8274-158X]{Andrew P. Cooper}
\affiliation{Institute of Astronomy and Department of Physics, National Tsing Hua University, Hsinchu 30013, Taiwan}
\affiliation{Center for Informatics and Computation in Astronomy, National Tsing Hua University, Hsinchu 30013, Taiwan}

\author[0000-0001-9667-1340]{Robert J.~J.~Grand}
\affiliation{Astrophysics Research Institute, Liverpool John Moores University, 146 Brownlow Hill, Liverpool, L3 5RF, UK}

\author[0000-0002-1947-333X]{Facundo A.~Gómez}
\affiliation{Departamento de Astronom\'ia, Universidad de La Serena, Av. Ra\'ul Bitr\'an 1305, La Serena, Chile}

\author[0000-0003-2325-9616]{Antonela Monachesi}
\affiliation{Departamento de Astronom\'ia, Universidad de La Serena, Av. Ra\'ul Bitr\'an 1305, La Serena, Chile}



\begin{abstract}
Ongoing large stellar spectroscopic surveys of the Milky Way seek to reconstruct the major events in the assembly history of the Galaxy. Chemical and kinematic observations can be used to separate the contributions of different progenitor galaxies to the present-day stellar halo. Here, we compute the number of progenitors that contribute to the \replyca{accreted} stellar halos of simulated Milky Way–like galaxies as a function of radius (the radial diversity) in three suites of models: Bullock \& Johnston, Aquarius, and Auriga. We show that there are significant differences between the predictions of these three models, beyond the halo-to-halo scatter expected in $\Lambda$CDM. Predictions of diversity from numerical simulations are sensitive to model-dependent assumptions regarding the efficiency of star formation in dwarf galaxies. We compare, at face value, to current constraints on the radial diversity of the Milky Way's \replyca{accreted halo}. \replyca{These constraints imply} that the halo of our Galaxy \replyca{is} dominated by \replyca{$\sim2$} progenitors in the range 8–45 kpc, \replyca{in contrast to averages of $7$ progenitors in the Bullock \& Johnston models, $3.5$ in Aquarius, and $4.2$ in Auriga over the same region}. We additionally find that the models with radial diversity most similar to that of the Milky Way are predominantly those with ongoing merger events. The Milky Way therefore appears unusual in having an \replyca{accreted} stellar halo dominated by a small number of progenitors accreted at very early times.

\end{abstract}




\section{Introduction} \label{sec:intro}
 
The tidal disruption of dwarf galaxies is \replyca{thought to be} the most important process responsible for building up diffuse, metal-poor stellar halos around galaxies like the Milky Way \citep{Searle:1978aa,Johnston:1996aa}. Clear evidence for this process is provided by the many stellar halo substructures that have been discovered in the Milky Way itself \citep[e.g.][]{Helmi:2020tx,Bonaca:2024aa}, around M31 \citep[e.g.][]{Ibata:2014}, and around other nearby galaxies \citep[e.g.][]{Martinez-Delgado:2010aa}. Some of these are associated with ongoing disruption of a surviving remnant, most notably the Sagittarius stream \citep{Ibata:1995aa,Newberg:2002aa,Majewski:2003aa}. The Gaia satellite has enabled rapid progress in understanding the number, nature, and origin of the progenitor galaxies that have contributed to the Milky Way's stellar halo \citep{Helmi:2020tx,Deason:2024aa}. In particular, many prominent structures have been discovered in phase space that were not readily apparent in configuration space, including the Gaia Sausage–Enceladus feature \citep[\gse{},][]{Haywood:2018ab,Helmi:2018aa,belokurov:2018_dischalo}. These heavily phase-mixed debris components likely correspond to the most massive and earliest accretion events \citep{belokurov:2018_dischalo}, which are expected to be relatively centrally concentrated in the Galactic potential 
\citep[e.g.][]{Johnston:2008aa,Amorisco:2017aa,Orkney:2022aa}. They can be identified as overdensities of stars with similar chemical abundances and orbital properties such as energy, angular momentum, or actions \citep[e.g.][]{Helmi:2000aa,McMillan2008,Gomez:2010ab, Nissen:2010aa,Myeong:2018aa}. Massive progenitors can also be identified by similar chemodynamical analyses of globular clusters \citep{Lynden-Bell:1995aa, Myeong:2018ac, Kruijssen:2020aa}. Chemical abundance and radial velocity measurements from ongoing deep and wide spectroscopic surveys will likely lead to more discoveries and better characterization of features already known \citep{Conroy:2019tc, 4most,desimws,Weave}. Nevertheless, current data are already sufficient to make quantitative estimates of the fraction of mass contributed to the stellar halo by its few most massive progenitors \citep[e.g.][]{H3,Naidu:2021aa}, albeit still with substantial uncertainty around how those progenitors are identified \citep[e.g.][]{Lane:2023aa,Carrillo:2024aa} and how local observations are extrapolated \citep[e.g.][]{Sharpe:2024aa}. These successes and challenges motivate direct comparison between the available data and the predictions of simulations.

In general, cosmological simulations of stellar halos associated with Milky Way–like galaxies appear to be in reasonable qualitative agreement with the observed mass, density, chemical composition, and kinematics of the real Milky Way stellar halo \citep{Bullock:2000aa, B&J2005,Cooper2010,Deason2016,Monachesi:2016ab, Monachesi2019,Font:2020aa}. Agreement in this context means that the (often highly uncertain) observed properties lie within the (considerable) system-to-system and model-to-model scatter predicted by the simulations for isolated systems of present-day virial mass \replyca{$0.5 \lesssim M_\mathrm{vir} \lesssim 2\,\times 10^{12}\,\msun$}. Different simulations also agree on the most significant factors that determine the bulk properties of stellar halos, including the efficiency of galaxy formation in low-mass progenitor at high redshift, as well as the effectiveness of tidal disruption (dependent on the structure both of the progenitors and the central galaxy). Simulations may also produce a halo-like component through in situ star formation, unrelated (or only indirectly related) to galactic accretion \citep{Abadi:2003aa,Tissera2013,Cooper:2015aa,Monachesi2019,Font:2020aa}. The presence of an `in situ' stellar halo is perhaps the most significant source of potential `confusion' in attempts to reconstruct the assembly history of the Milky Way \citep[e.g.][]{Rey:2022aa}. \replyjn{Chemical and kinematic evidence points to an in situ contribution to the Milky Way halo, at least in the inner galaxy, in particular from stars formed in the proto-Milky Way and later scattered onto eccentric orbits \citep[e.g.][]{Nissen:2010aa, Gallart:2019,Belokurov:2020}.}

\apc{Prior work on the composition of stellar halos has established that
a small number of the most massive disrupted progenitors most likely account for the bulk of the total stellar halo mass. We discuss this more quantitatively in section~\ref{sec:define_diversity}. In a very broad sense}, this a readily predictable consequence of the steep decline of star formation efficiency toward lower virial mass, combined with the characteristic accretion histories of Milky Way analogs \citep{Bullock:2000aa,Purcell:2007aa,B&J2005,Cooper2010,Cooper:2013aa,Deason:2013_broken,Deason2016,Monachesi2019,DSouza:2018aa}. 

\replyjn{A range of scenarios for the assembly of the accreted stellar halo arise from the range of CDM halo mass growth histories at a fixed present-day virial mass \citep[e.g.][]{Amorisco:2017ac}. The picture is more complicated than might be expected based on predictions for satellite merger rates alone, because satellites must be heavily disrupted in order to become significant stellar halo progenitors. On average, satellites with the highest total stellar mass arrive late, and are less likely to be disrupted by $z=0$. For a Milky Way analog, the dominant stellar halo progenitors are therefore typically less massive than the most massive surviving satellite, and early-arriving massive progenitors can make relatively significant contributions. This suggests a complex but potentially informative relationship between the assembly history of a galaxy and the mass, diversity, and structure of its accreted stellar halo. For example, \citet{Amorisco:2017ac} show that, on average, the total mass of the accreted stellar halo is maximized when a relatively larger number of progenitors of moderate mass ($M_\star \sim 10^{8}\,\msun$) arrive at intermediate redshift ($z\sim1.5$). This results in a high stellar mass per progenitor and allows enough time to incorporate those satellites into the stellar halo. At fixed present-day virial mass, such an assembly history requires the system to lag behind the average growth rate at earlier and later times.}

\replyjn{These expectations from simplified models are reflected in cosmological simulations, although the stochastic nature of massive accretion events can obscure trends in small samples. In general, simulated accreted halos of Milky Way analogs typically comprise a few substantial contributions from mergers at early times, along with a further one or two massive accretions of similar or greater stellar mass at late times \citep[e.g.][]{B&J2005, Cooper2010,DSouza:2021aa,Horta:2024aa}.}

There is some evidence suggesting that the Milky Way may be an example of \replyjn{a system assembled rapidly and early,} in which a small number of early-\replyjn{arriving} \replyjn{moderately} massive progenitors dominate the total \replyjn{accreted} stellar halo mass \citep[e.g.][]{Deason:2013_broken}. \replyjn{However, the bulk of the mass from such progenitors} is expected to be deposited \replyca{close to} the center of the halo \replyjn{\citep[e.g.][]{Amorisco:2017aa}}, largely out of reach of current \replyca{surveys of the Milky Way}. 

\replyjn{Observations of Milky Way analogs are also in line with the expectation that $\lesssim10$ progenitors account for the bulk of the accreted mass, with a large fraction contributed by the most massive progenitor.}
In particular, where individual halo stars can be resolved, the tight galactic mass-metallicity relation enables the total stellar mass of the most massive progenitor to be estimated (to an accuracy of $\lesssim 1$~dex) from the peak of the metallicity distribution function \citep[e.g.][]{Harmsen:2017aa, Bell:2017aa, Jang:2020ab, Gozman:2023aa}. This can be combined with an estimate of the total accreted mass (e.g.,\ from fits to the outer regions of the surface brightness profile) to determine the extent to which the most massive progenitor dominates; this approach has been validated in cosmological simulations \citep[e.g.,][]{DSouza:2018aa,Monachesi2019}.
Using the Auriga simulations, \citet{Monachesi2019} find that the radial gradient of metallicity can, in some cases, further constrain the number of less massive progenitors that make significant contributions to the halo.

\replyca{The generic expectation described above -- that a small number of progenitors will dominate the accreted stellar halo at the virial mass scale of the Milky Way -- relates to the entirety of the accreted mass. In practice, in particular in the Milky Way, observations can presently only sample a small fraction of the total volume of the halo, which may lead to different conclusions about the nature and number of the most} significant progenitors \citep[e.g.][]{Sharpe:2024aa}. The relative \replyca{significance of different progenitors may vary} with radius \citep{Cooper2010,Deason:2013_broken,Deason:2014aa, Deason2016, Amorisco:2017aa, Fattahi:2020aa}. It is therefore relevant to ask if simulations predict that the region of the \replyca{accreted} stellar halo probed by current surveys \replyca{of the Milky Way} (i.e.,\ within heliocentric distances of $\lesssim20$~kpc) is dominated by one object or many, and whether the inferred masses and accretion times of the significant progenitors inferred from the Milky Way data are consistent with expectations for a `typical' $10^{12}\,\msun$ Milky Way host.

Our aim in this paper is to explore the radial dependence of the `diversity' of progenitor contributions in Milky Way analog stellar halos using three commonly cited suites of Milky Way stellar halo simulations: the semi-analytic/hybrid $N$-body model of \citet{B&J2005}, the semi-analytic particle-tagging Aquarius models of \citet[][]{Cooper2010} and the hydrodynamical Auriga models of \citet{Grand2017}. We compare these predictions, at face value, with the inferences made from the current data by \citet{H3}, which are broadly representative of recent interpretations of Milky Way observations. The simulations differ in their numerical approach and assumptions regarding the physics of galaxy formation. Our goal is to summarize the systematic differences between their predictions, in the specific contexts in which they are most readily compared to current stellar halo observations. To support our comparison, we also make available the underlying simulation data of \citet{Cooper2010}.

In detail, this kind of comparison is best conducted with mock observables \citep[e.g.][]{Lowing:2015aa,Grand:2018ab,Sanderson:2018aa,Kizhuprakkat:2024ab}, rather than directly from the `raw' simulation data, in order to account for observational biases and systematic uncertainties, for example in total densities estimated from small samples of tracers. Each of the three simulations we examine has been used to make publicly available mock star catalogs, which could be used for this purpose. Here, however, we use the raw simulations because our aim is only to establish a `baseline', for the range of underlying predictions from the models, independent of the parameters and selection function of any particular survey. We hope this will be useful to underpin more detailed future comparisons between specific surveys and mock catalogs.

The paper is structured as follows. In Section \ref{sec:simulation}, we briefly summarize the assumptions, methods, initial conditions, and relevant results from each of the three simulation suites (Sections \ref{sec:bj05} to \ref{sec:auriga}). We compare their predictions for \replyca{accreted} stellar halo masses and density profiles in section \ref{sec:accreted_def}. In Section \ref{sec:diversity} we present their radial diversity profiles. We summarize our findings and conclude in Section \ref{sec:conclusions}.

\begin{deluxetable*}{lcccccccccc}
     \label{tab:1}
     \tabletypesize{\scriptsize}
     \tablewidth{0pt}
     \tablecaption{Parameters of the BJ05, Aquarius, and Auriga simulation suites. From left to right, columns show representative \replyca{dark matter particle masses (for Auriga, the star particle mass is given in parentheses)}, the gravitational force softening lengths \citep[physical maximum softening at $z=0$; see, e.g.,][]{Grand:2024aa}, virial mass range of the central dark matter halos, number of halos in each suite ($N_{\text{halo}}$) and cosmological parameters (Hubble parameter, density parameters of all matter, baryonic matter and the cosmological constant, as well as the present-day matter power spectrum normalization, $\sigma_{8}$).}
     \tablehead{
     \colhead{ } & \colhead{Numerical Method} & \colhead{$N_\mathrm{halo}$} & \colhead{ Particle Mass} & \colhead{Softening Lengths} & \colhead{Halo Masses} & \colhead{$h$} & \colhead{$\Omega_{m}$} & \colhead{$\Omega_{b}$} & \colhead{$\Omega_{\Lambda}$} & \colhead{$\sigma_8$}\\
    \colhead{ } & \colhead{} & \colhead{ } & \colhead{($10^{3} \mathrm{M}_\odot$)} & \colhead{(pc)} & \colhead{($10^{12} \text{M}_\odot$)} & \colhead{ } & \colhead{ } & \colhead{ } & \colhead{ } & \colhead{ } 
     }
     \startdata
     BJ05 & $N$-body & 11 & - & - & 1.4 & 0.7 & 0.3 &0.048 & 0.7 & 0.9\\
     Aquarius & $N$-body & 6 & $\sim10$ & 65.75 & 0.82 - 1.84 & 0.73 & 0.25 & 0 & 0.75 & 0.9\\
     Auriga L4 & MHD & 30 & $\sim300$ [50] & 375 & 0.92 - 1.91 & 0.6777 & 0.307 & 0.048 & 0.693 & 0.8288\\
     \enddata
 \end{deluxetable*}

\section{Simulations} 
\label{sec:simulation}

In this section, we briefly describe the three sets of simulations \replyca{and compare their predictions for accreted stellar halos}. Table~\ref{tab:1} lists the cosmological parameters assumed by each model. Each model assumes a slightly different cosmology, but we do not attempt to homogenize them in our comparison. In particular, where the simulations provide quantities scaled to $H_{0} = 100\,h\,\mathrm{km\,s^{-1}\,Mpc^{-1}}$, we convert these to `observed' quantities using the value of $h$ from the cosmological parameter set assumed by the simulation (i.e. we use different values of $h$ for each simulation, as in Table~\ref{tab:1}).
 
\subsection{Bullock \& Johnston simulations}
\label{sec:bj05}

The \citet[][hereafter BJ05]{B&J2005} models comprise 11 realizations of \replyca{accreted} stellar halos in dark matter halos with present-day virial mass $M_{\text{vir,0}} = 1.4\times10^{12}\,\msun$ (corresponding to a virial radius $R_{\text{vir,0}}= 282\,\mathrm{kpc}$ and a virial velocity $V_{\text{vir}} = 144\,\mathrm{km\,s}^{-1}$, in their cosmology). A Monte Carlo method based on the Extended Press–Schechter formalism was used to generate the masses and infall times of the dwarf galaxy progenitors in each realization. A separate $N$-body simulation was then carried out for each individual progenitor, running from the time of infall to the present day. The central galaxy was represented by a rigid analytic potential, combining contributions from a disk, a bulge, and a spherical dark halo with a Hernquist density profile. The parameters of the potential were varied smoothly with time, to represent the mass and size growth of the disk and halo. In each individual $N$-body simulation, the infalling progenitor was represented by $10^{5}$ equal-mass particles with initial conditions (concentration and orbital parameters) drawn from parameterized distributions. Dynamical friction from the host halo was included by imposing an analytic `drag' force on the progenitor particles. The stellar mass associated with a progenitor was represented by tagging its particles with energy-dependent stellar mass weights, assigned in such a way as to make the `stellar' energy distribution function in the initial conditions that of a Plummer sphere. The corresponding stellar populations were assigned using empirical distributions derived from observed properties of Milky Way satellites. 

BJ05 list several caveats of their approach. The star formation histories of satellites are greatly simplified. The analytic potential does not account for the response of the central galaxy to its satellites. The method also neglects the possibility of interactions between satellites, which may be stronger at earlier times or in halos that have assembled relatively recently from smaller groups of similar mass. BJ05 suggest these limitations are likely to be important only for early accretion events (prior to the last major merger). Since such events are predominantly associated with the assembly of the inner halo, they state that their model most reliably predicts the spatial and velocity structure of accreted stars at distances $\gtrsim 20$ kpc. Taking this limitation into account, the 11 accretion history realizations in BJ05 were selected such that no more than 10 percent of the virial mass of the host was accreted in the last 7~Gyr before the present day in each case.

Notwithstanding these limitations, the BJ05 models \citep[supplemented by][]{Robertson:2005aa,Font:2006aa} have long served as the most widely used and influential set of numerical predictions for the phenomenology of galactic stellar halos, and for interpreting observations of the Milky Way and M31 \citep[e.g.,][]{Bell:2008,  Johnston:2008aa,sharma:2011_galaxia,Ibata:2014,Sharpe:2024aa}. Among their many results, the BJ05 models were the first to clearly demonstrate the principle discussed in the introduction, i.e., that the bulk of the mass in the stellar halo of a typical Milky Way analog can be (and often is) deposited by a relatively small number of satellite progenitors with total mass $M_\mathrm{vir}\sim10^{10}\,\msun$. BJ05 quantify this by the fraction of \replyca{accreted} stellar halo mass associated with the 15 most massive progenitors, which they find to be $\gtrsim80~\%$ in all 11 halos (their Table~1). \citet{Robertson:2005aa} and \citet{Font:2006aa} established that the chemical abundances observed for Milky Way halo stars further support growth from a small number of massive progenitors, and moreover, that those progenitors must have been accreted relatively early, $>9$ Gyr ago. 

 \subsection{Aquarius}
 \label{sec:aquarius}
 
 Aquarius is a suite of six high-resolution cosmological $N$-body `zoom' resimulations of Milky Way–like dark matter halos \citep{Springel2010}. \citet[][hereafter C10]{Cooper2010} processed merger trees for the `level 2' Aquarius simulations with the semi-analytic galaxy formation model \textsc{galform} \citep{Cole2000,Bower:2006aa,Font:2011aa}. The star formation histories predicted by \textsc{galform} were used as the basis for a particle-tagging technique, which associates single stellar populations with subsets of tightly bound dark matter particles in a `continuous' fashion along each branch of the merger tree. Using the tagged particles, the evolving three-dimensional spatial distribution and kinematics of each stellar population can be tracked through the $N$-body simulation in post-processing, under the assumption that the baryonic mass does not significantly perturb the potential as represented by the underlying $N$-body model. This procedure is described in detail by \citet{Cooper2017}, along with a study of its limitations and correspondence to the results of hydrodynamical methods. The properties of each Aquarius stellar halo from C10 are summarized in the Appendix~\ref{Appendix1}. With this paper, we make available the particle data from the C10 models (see the data availability statement).
 
 In C10, in contrast to BJ05, the evolution in the mass and structure of the main halo and its satellites are calculated \textit{ab initio}; in this $N$-body sense, the Aquarius simulations are `fully cosmological' and dynamically self-consistent. The central potential is nonspherical, and it can grow violently as well as through smooth accretion. \replyca{Aquarius is therefore a more realistic model than BJ05 at early times, prior to the last merger that strongly perturbs the central potential. Moreover, the Aquarius sample is not restricted to particular assembly histories, and hence, unlike BJ05, it includes halos that grow significantly through the accretion of relatively massive progenitors at $z<1$.} Nevertheless, the particle-tagging scheme involves significant approximations. In particular, it does not account for the growth of a massive baryonic disk in the main halo, which may have a substantial effect on the disruption of satellites, both those accreted at early times and massive satellites accreted relatively late \citep{Cooper2017}.

 In the Aquarius models, the separation between `in situ' and `accreted' components of the central galaxy is straightforward (see also below). In situ stars are defined as those that form in the main branch of the merger tree (these were ignored entirely in C10; we ignore them here as well). The accreted component comprises all other stars bound to the central galaxy at $z=0$ (as opposed to those still bound to surviving satellite galaxies).

 C10 further separated the accreted component into two regions: `bulge' (galactic radius $R<3\,\mathrm{kpc}$) and `stellar halo' ($3<R<200\,\mathrm{kpc}$). Here, we study the properties of accreted stars as a function of radius, so we treat the $R<3\,\mathrm{kpc}$ region in the same way as the rest of the accreted component.

 As shown in \citet{Cooper2010}, the six Aquarius halos have very different accretion histories, reflected by differences in the present-day distribution and metallicity of their accreted stellar halos. One reason for this variation is that the Aquarius halos span a factor-of-two range in virial mass, corresponding roughly to the scatter in estimates of the true MW virial mass \citep{Wang:2020aa}. This contrasts with the BJ05 models, all of which have identical virial mass. However, the small size of the Aquarius sample obscures any trend with virial mass. Instead, the variations from halo to halo are dominated by stochastic differences in the timing and properties of the few most massive accretion events. Although these properties are correlated with virial mass on average, the observable properties of a single stellar halo depend very strongly on whether or not that most massive satellite has been disrupted or not. We return to this point in our short comparison of the different models below. 
 
 \subsection{Auriga}
 \label{sec:auriga}
 
 Auriga is a suite of $N$-body magnetohydrodynamical zoom simulations of Milky Way–like galaxies from cosmological initial conditions \citep{Marinacci2014, Grand2017, Grand:2024aa}. The simulations were carried out with the \texttt{AREPO} code \citep{Springel2010}. The original\footnote{\replyca{The recent Auriga data release described in \citet{Grand:2024aa} includes nine additional halos with lower virial masses, $10^{11}--10^{12}\,\mathrm{\msun}$. These are not considered here or in earlier papers about stellar halos in Auriga.}} Auriga suite comprises 30 halos simulated at a baseline resolution (level 4, L4), labeled Au01 to Au30. Of these, six were re-simulated at even higher resolution (level 3, L3). We use only the L4 halos in this work. Our study is concerned with the bulk structure of the halos, which for the most part is dominated by the most massive accreted progenitors. The lower resolution of L4 is therefore not as significant as the benefit of a much larger sample. 
 
 In contrast to Aquarius and BJ05, Auriga includes star formation and the dynamical evolution of gas and stars self-consistently. This is particularly important for modeling the growth of the central galaxy and its response to individual accretion events. The distribution of orbits for halo stars should therefore better reflect both the phase space distribution of stars in satellites before they are disrupted, as well as interactions between the central galaxy and the accreted satellites (including stronger dynamical friction, more rapid disruption, and tidal shocking). In principle, a hydrodynamical treatment also allows for an `in situ halo' formed \replyca{by stars `kicked' out of the disk during mergers} or in gas clouds stripped from satellites, before they have time to mix into the gaseous halo or interstellar medium of the central galaxy \citep[e.g.][]{Cooper:2015aa}. However, as discussed by \citet{Cooper2017}, it is not obvious that \replyca{these} effects will dominate the differences between particle tagging and hydrodynamical realizations of halos simulated from the same initial conditions: such differences are more likely to be dominated by model-to-model variations in the star formation histories of individual satellites predicted by different combinations of hydrodynamical and subgrid methods. 
 
 The Auriga galaxies were chosen from a lower-resolution cosmological volume, based on a set of criteria to identify Milky Way analogs \replyca{on the basis of their virial mass and isolation} \citep[see][]{Grand2017,Grand:2024aa}. The resulting galaxies have radial scale lengths ranging from $2.16$ to $11.64\,\mathrm{kpc}$, most with prominent central bulge components. \replyca{As described in \citet{Monachesi2019}}, their stellar halos, \replyca{defined kinematically by stellar particles with circularity parameter $\epsilon < 0.7$ and radius $> 5$~kpc (to avoid the bulge region)}, have masses in the range $10^{9}$--$10^{10}~\rm M_{\odot}$, with a significant contribution from stars formed in situ. \replyca{The Auriga halos have a wide range of properties, consistent with the Aquarius halos in C10}. Previous studies of Auriga have \replyca{the above kinematic definition of the stellar halo, as well as other definitions}; for consistency with the BJ05 and Aquarius models, we use a simple merger tree-based definition in this paper, described below.
 
The Auriga halos have been shown to contain phase-space structures resembling the Milky Way's \gse{} feature. \replyca{Among the 30 L4 halos, \citet[][see their Figure~3]{Fattahi2019} identified 10  with velocity anisotropy and mass fraction resembling \gse{} (they excluded Au11 and Au20 from their study, because those halos undergo low mass ratio mergers close to $z=0$). Among those, four (Au05, Au09, Au10, and Au18) were identified as particularly close analogs. \citet{Orkney:2023aa} further investigate the properties of these potentially \gse{}-like progenitors, finding that they typically have pre-infall virial masses $\sim3$--$11\times10^{10}\,\mathrm{\msun}$, high gas fractions $40$--$80\%$ and a wide range of infall times, $0.9<z<2.9$. \citet{Orkney:2023aa} identify a handful of cases (e.g.,~Au10) in which more than one progenitor contributes to the \gse{}-like debris. We consider this latter point in section~\ref{sec:compare_mw}.}

 \subsection{Definition of the accreted stellar halo and its progenitors}
 \label{sec:accreted_def}

All the models focus on an isolated Milky Way analog at $z=0$. We refer to the dark matter halo of this galaxy as the \textit{main halo}. We define the \textit{accreted stellar halo} of the galaxy to be the set of simulated star particles that \textit{were not formed in the main branch} of the merger tree of the main halo, as described below.

By construction, all the stellar mass in the BJ05 models is accreted according to this definition. For Auriga \replyca{and Aquarius (see also \ref{sec:aquarius})}, we use merger trees built as described in \citet{Jiang:2014aa} to define the main branch as the chain of most-massive progenitors of the main halo, based on the identification of self-bound dark matter subhalos with the SUBFIND algorithm \citep{Springel:2001aa}. We trace the set of star particles bound to the main halo at $z=0$ back in time. We classify a particle as \textit{accreted} if, at an earlier simulation snapshot, it is bound to a halo other than the main branch progenitor of the main halo. The snapshot at which the particle is last bound to another branch of the merger tree defines its \textit{accretion time}. We classify particles that are always bound to the main branch as \textit{in situ}. We assign a common \textit{progenitor ID} to all accreted particles associated with a single progenitor branch. The progenitor ID is the basis for most of our analysis\footnote{\replyca{Our progenitor ID was computed independently of the accreted progenitor label recently made available by \citet{Grand:2024aa}, although the two labels are likely to be very similar.}}. 

Particles formed in long-lived satellites of the main halo may not be accreted into the stellar halo of the Milky Way analog until long after their parent halo has crossed the virial radius of the main branch (the \textit{infall time} of the progenitor). A given progenitor ID therefore corresponds to a single infall time, but it may be associated with a wide range of individual particle accretion times. Satellite halos may continue to form stars after they become subhalos; if such stars are unbound from their parent satellite at $z=0$, we label them as accreted, consistent with the treatment of stars formed in the same progenitor before it became a subhalo of the main branch. 

\replyjn{Since we provide progenitor IDs as part of the datasets that accompany this paper, we remark on the following somewhat subtle detail. The} progenitor ID indicates the branch \textit{from which a star was unbound} when it became part of the stellar halo of the Milky Way analog: this is \textit{not} necessarily the branch in which the star \textit{formed}, because each subhalo joining the main branch has its own set of hierarchical progenitor branches. In the case of the most massive satellites, more than one of these `satellite of satellite' branches may form stars. \replyjn{For example, say that galaxy B is a satellite of A, which itself becomes a satellite of the Milky Way analog. When this pair of galaxies crosses the virial radius of the Milky Way analog, B may be separated from A by tidal forces. Stars subsequently stripped from B will then be accreted directly into the stellar halo of the Milky Way analog. Our algorithm will assign a distinct progenitor ID to those stars (i.e.,\ for those stars, B will be treated as a direct progenitor of the MW analog). However, stars from B that are first accreted by A (before or after A becomes a satellite of the MW analog) will instead be labeled with the progenitor ID of galaxy A. Consequently, it is possible for stars formed in one progenitor branch (the branch of galaxy B in this example) to} have different progenitor IDs. This is consistent with our intended purpose here and is likely rare in practice\footnote{Several technical subtleties are involved building merger trees by linking structures of halo-finding methods (like SUBFIND) across the snapshots of $N$-body simulations. For example, during complex merger events, the operational definition of the main branch as the most massive progenitor can be ambiguous, and links between `related' branches or sections of branches can be broken \citep[see][for extended discussion of the potential problems]{Jiang:2014aa}.  The most significant impact of such uncertainties on our work is a small number of stellar particles that cannot be ascribed to any accretion event. We have not observed any cases in which these problematic particles contribute significant mass to the \replyca{accreted} stellar halo.}.

In hydrodynamical simulations like Auriga, in situ stars can form directly on halo-like orbits \replyca{or be scattered onto such orbits by gravitational interactions, after having formed in the central disk} \citep[e.g.][]{Abadi:2003aa,Zolotov:2009aa, Cooper:2015aa,Font:2020aa}. Alternative definitions of the stellar halo may count some fraction of these as `halo' stars in order to make a more realistic comparison with observations (for example, selecting stars on very eccentric orbits). We prefer \replyca{an idealized separation of accreted and in situ stars, using knowledge of the merger history from the simulations,} because our purpose is mainly to compare different models \replyjn{(Auriga may overpredict the mass of in situ stars on halo orbits, which complicates any comparison that mixes accreted and in situ halo stars; \citealt{Monachesi2019})}. In future work, it would be useful to examine how in situ stars affect inferences about the composition of the accreted stellar halo.

\subsection{Comparison between models}

 \subsubsection{Stellar halo masses}
\label{sec:modelcomparison}

 \begin{figure}
    \centering
    \includegraphics[width = \columnwidth]{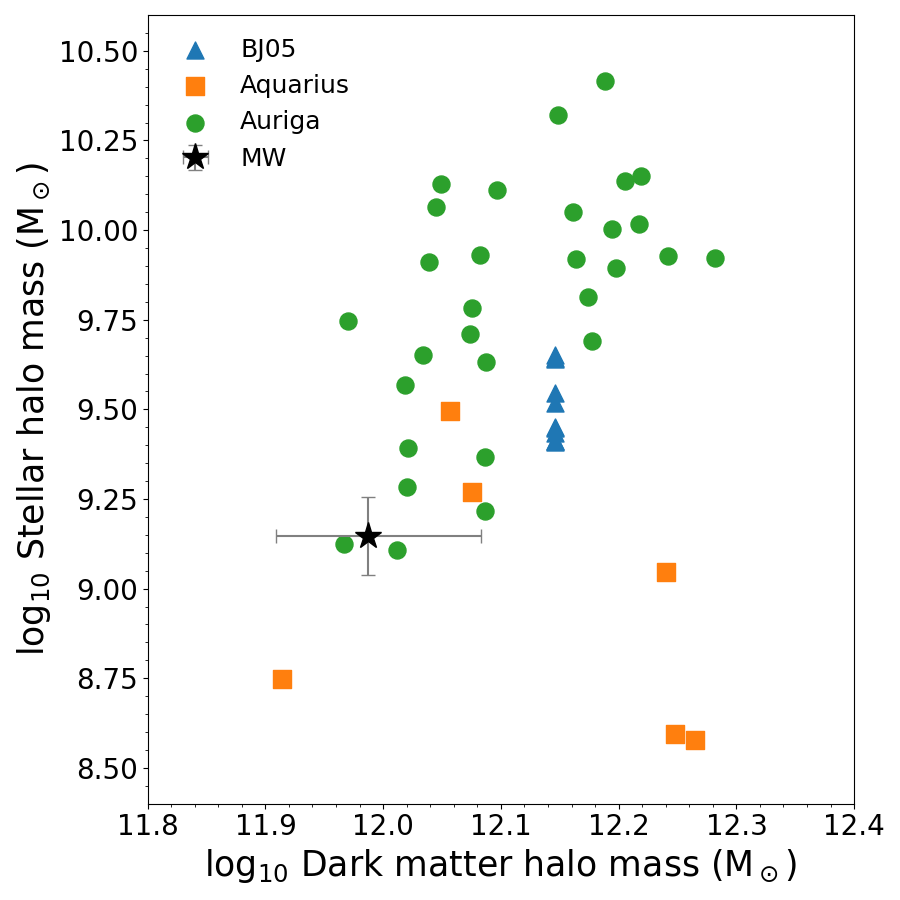}
    \caption{Virial masses and \replyca{accreted} stellar halo masses of galaxies in the three model sets we use in this work (symbols, as shown in the legend). \replyjn{The star symbol shows recent estimates for the Milky Way (stellar halo mass $M_{\star}=1.4\pm0.4\times10^{9}\,\msun$ \citep[][see also \citealt{Mackereth:2020aa}]{Deason:2019aa}; virial mass $M_{200}=0.97^{+0.24}_{-0.19}\times10^{12}\,\msun$ (\citealt{Cautun2020}; see also \citealt{Wang:2020aa}})).}
    \label{fig:1}
 \end{figure}
 
 Figure~\ref{fig:1} shows the virial mass and the total accreted stellar halo mass for each halo. The star symbol shows currently favored values for the Milky Way \citep{Licquia:2015ww,Wang:2020aa}. The models span slightly more than a factor of two in virial mass, with most systems more massive than the Milky Way. Accreted stellar halo masses range from \replyjn{\replyca{$10^{8.5}$} to $10^{10.5}\,\msun$} (in most cases, significantly larger than the Milky Way's halo mass). The Auriga simulations account for most of this scatter, although this may be simply because of their larger sample size. However, there are some clear differences between the three sets of models. Auriga shows a weak trend in \replyca{accreted} stellar halo mass with virial mass. This is an expected consequence of the overall relationship between total stellar mass and virial mass \citep{Purcell:2007aa,Cooper:2013aa}. \replyjn{The typical stellar mass at fixed (peak) virial mass in Aquarius is somewhat less than that in Auriga (see Figure~\ref{fig:SMHMrelation}), leading to a lower total mass for the dominant progenitors of the accreted stellar halo in Aquarius. The apparently shallower trend of accreted stellar halo mass with virial mass in Aquarius is harder to interpret as a systematic difference with Auriga here, because of the small number of Aquarius halos and the large scatter among the Auriga galaxies.}
 %
 %
 In the case of BJ05, the system-to-system scatter (which is due only to variations in star formation and accretion history at fixed virial mass) is clearly much smaller than that predicted by Auriga. This may be a consequence of the restricted range of accretion histories in BJ05, which excluded recent mergers with massive satellites.
 
 As we discuss below, despite this large system-to-system scatter, the diversity of individual stellar halos shows clear systematic differences between the three model sets. We note that Auriga provides the simulations that are closest to measurements of the virial mass and stellar halo mass of the Milky Way (e.g. Au09, Au10, Au17, and Au22).

\subsubsection{Progenitor masses}

  \begin{figure}
     \centering
     \includegraphics[width=\columnwidth]{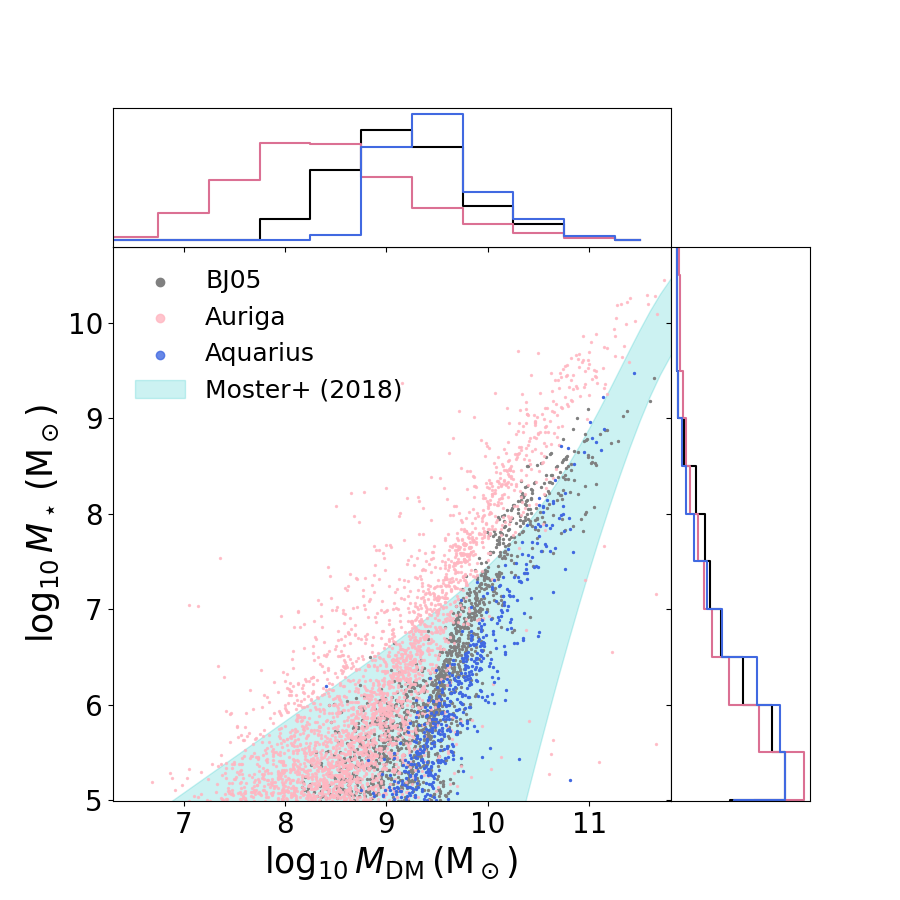}
     \caption{The relation between stellar mass and dark matter halo mass (prior to accretion) for progenitor satellites in the \replyjn{6 Aquarius,} 30 Auriga, and 11 BJ05 models. Progenitors from all individual halos in each simulation set are combined in this plot. The top and right-hand panels show the normalized distributions of halo mass and stellar mass, respectively.}
     \label{fig:SMHMrelation}
 \end{figure}

Figure~\ref{fig:SMHMrelation} shows the relationship between stellar mass and dark halo mass (approximately virial mass) for all progenitor satellites in \replyjn{Aquarius (blue),} Auriga (pink), and BJ05 (gray). The total virial mass here is defined prior to infall of the progenitor to the main halo, hence this plot shows the `underlying' relation, without the effects of tidal stripping on surviving satellites. The figure provides some insight into the different star formation efficiencies for satellites predicted by the models. They predict almost identical luminosity function shapes (right-hand histogram). Even well-resolved satellites in Auriga occupy somewhat less massive halos than satellites of comparable \replyca{stellar} mass in BJ05 or Aquarius (this can be seen in the shift of the halo mass distribution to lower halo mass for Auriga progenitors, as shown in the top histogram). The joint distributions show similar overall trends, but it is clear that Auriga progenitors have lower virial mass at fixed stellar mass (or equivalently,  higher stellar mass at fixed virial mass)\replyjn{, while BJ05 and Aquarius occupy almost the same region.} The difference is greatest (up to $\sim1$ order of magnitude) for the most massive satellites. 

The most significant consequence of these differences is that the most massive progenitors in Auriga can contribute relatively more mass to the \replyca{accreted} stellar halo. Together with other effects such as dynamical friction, this is likely to lead to systematic differences between the diversity profiles of the models, as we discuss further below.

\subsubsection{Overall density profiles}
\label{sec:density_profiles}

\begin{figure}
    \centering
    \includegraphics[width=\columnwidth]{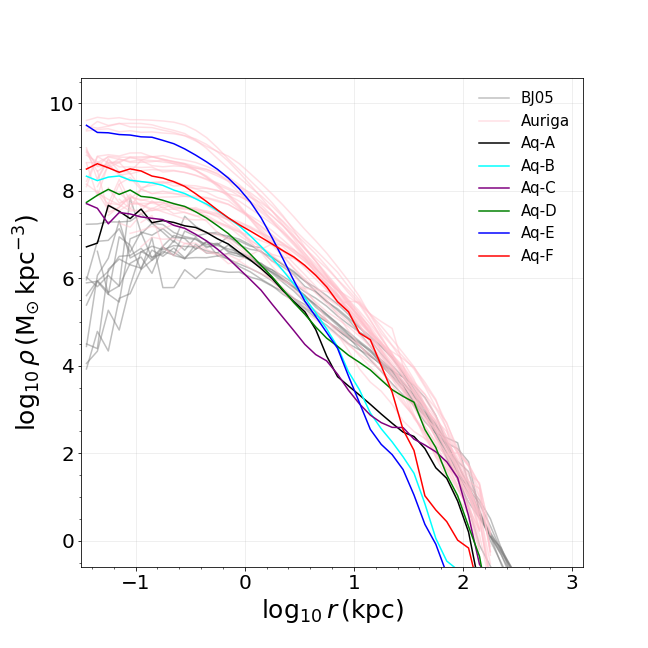}
    \caption{A comparison of the volume density profiles of the accreted stellar halos in each simulation. The six Aquarius halos are shown with individual colors. For simplicity, the 11 BJ05 halos and 30 Auriga halos are shown with gray and pink lines, respectively.}
    \label{fig:2}
 \end{figure}

Figure~\ref{fig:2} shows the overall distribution of accreted stars in the different models. The radial scales and outer power-law slopes are broadly consistent among the three models, but there are also clear systematic differences. The Auriga halos have the highest average density at all radii and are notably similar to one another in their outskirts, with $\lesssim 1~\mathrm{dex}$ variation in density between different galaxies at all radii $\gtrsim 10~\mathrm{kpc}$. The BJ05 halos have lower density than Auriga at radii $\lesssim 30~\mathrm{kpc}$ and even less galaxy-to-galaxy variation. The lower central densities of the BJ05 models (and perhaps also their lower average accreted stellar masses, at fixed virial mass) may be the result of the limitations of those models at higher redshift, as discussed above.

The Aquarius halos have lower density than most of the BJ05 and Auriga halos at radii $\gtrsim 10~\mathrm{kpc}$ and show greater variation from galaxy to galaxy, as well as a wider range of slopes and more pronounced inflections (`breaks'). These breaks correspond to transitions between regions dominated by different combinations of progenitors, as described in C10. In some (not necessarily all) cases, the breaks correspond approximately to the apocenters of stars from individual dominant progenitors, and a pronounced break in the overall profile may indicate the mass of the \replyca{accreted} stellar halo overall is dominated by concentrated debris from a single progenitor \citep[][]{Deason:2013_broken}. As we explore further below, the most significant difference between the Auriga and Aquarius stellar halos is that debris from individual massive satellites generally makes a dominant contribution to the overall density profile across a much wider range of radii in Auriga.

\section{The diversity of accreted halos}
\label{sec:diversity}

\subsection{Definition of diversity}
\label{sec:define_diversity}
 
 To quantify the extent to which the  \replyca{accreted} stellar halo at a particular radius is dominated by the contribution of one progenitor or many, we define the \textit{number of significant progenitors},
 \begin{equation}
     N_{\text{sig}}(r) = \Big[\sum_im_i(r)\Big]^2\Big/\sum_i\left[m_i(r)^2\right],
     \label{eq:1}
 \end{equation}
where $m_i(r)$ is the stellar mass contributed to the accreted halo in an interval of radius ($r, r+\Delta r$) by the $i$th progenitor. 
$N_{\text{sig}} \simeq 1$ where a single progenitor dominates, and $N_{\text{sig}}\simeq N$ if $N$ progenitors contribute equal fractions of the accreted mass. This quantity was used by \citet{Cooper2010} to describe the overall diversity of the Aquarius stellar halos, and also by \citet{Cooper2017}, where it was computed as a function of radius (as in equation \ref{eq:1}) for a single halo.
 
Other measures of \textit{overall} stellar halo diversity (i.e.,\ for the whole \replyca{accreted} stellar halo) have been used in the literature. For example, as noted above, BJ05 reported the fraction of \replyca{accreted} stellar mass contributed by the 15 most massive progenitors. \citet{Deason2016} reported the mass-weighted average progenitor stellar mass, which (as they explain) is directly related to our $N_\mathrm{sig}$. Another example is that of \citet{Monachesi2019}, who used a measure we call $N_{90}$, defined as the number of the progenitors required to account for 90\% of the accreted stellar halo mass, when accumulated in descending order of progenitor mass and rounded up to the nearest integer \apc{(a similar measure is discussed in C10)}. For example, a halo in which the most massive progenitor accounts for $80\%$ of the \replyca{accreted} stellar halo mass, and the second most massive progenitor for a further $15\%$, would have $N_{90}=2$. \replyjn{By construction, $N_\mathrm{90} \gtrsim N_\mathrm{sig}$.} When a few massive progenitors account for most of the mass, but not more 90\%, and the remaining mass up to 90\% is dominated by a relatively large number of much less massive progenitors, \apc{then $N_{90}$ exceeds $N_\mathrm{sig}$ by a larger factor}.

\begin{figure}
    \centering
    \includegraphics[width = \columnwidth]{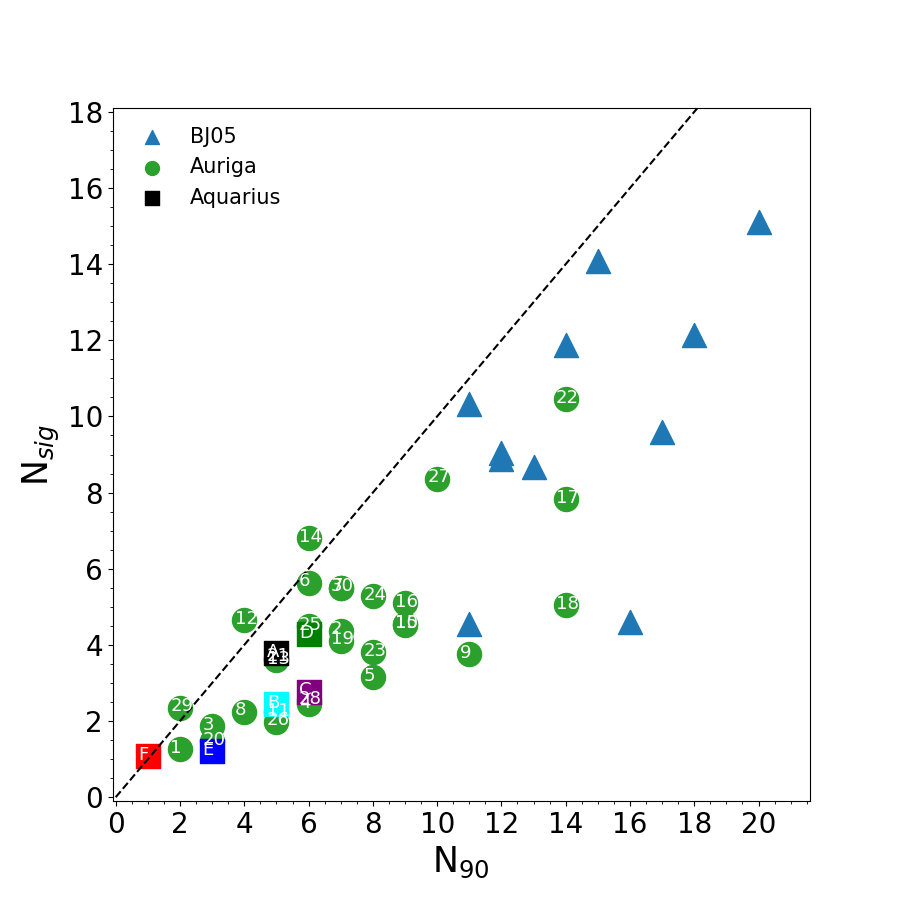}
    \caption{Our diversity measure \replyca{for the whole \replyca{accreted} stellar halo}, $N_\mathrm{sig}$, compared to the alternative measure $N_{90}$ \citep[e.g.][]{Monachesi2019}. Symbols correspond to model sets as shown in the legend, with labels showing the individual system numbers (Auriga) and letters (Aquarius). The color coding of the square symbols for Aquarius corresponds to that used in other figures and in \citet{Cooper2010}. The dashed line is the $1:1$ relation.}
    \label{fig:N90comparison}
 \end{figure}

Figure~\ref{fig:N90comparison} compares $N_{90}$ with $N_\mathrm{sig}$ for the halos in the three simulation sets. It is clear that the different models predict different distributions of progenitor mass for the accreted stellar halo. The BJ05 and Auriga models contain examples of systems in which \replyca{most of} the accreted mass is contributed by \replyca{many} progenitors, \replyca{each only a small fraction of the total} (high values of $N_\mathrm{sig}$ and \apc{$N_{90} > N_\mathrm{sig}$)}. Approximately 2/3 of the Auriga sample have $N_\mathrm{sig} > 3$; all but two of the BJ05 sample have $N_\mathrm{sig}>5$. \apc{Furthermore, the BJ05 has no halos with $N_\mathrm{sig} < 4$, and hence only limited overlap with the distribution of Aquarius and Auriga halos.}

\apc{This figure, for either measure of diversity, demonstrates that, when invoking the rule of thumb that `a few massive progenitors dominate a Milky Way–like halo', `a few' should be understood as `fewer than 10', rather than `one or two.' The simulations we use are consistent with the earlier literature in this respect} -- for example, \citet{Deason2016}, using an empirical method for assigning stellar masses to progenitors in $N$-body simulations of 45 halos with virial mass $10.2\times10^{10}\,\msun$, found that $\sim50\%$ of their sample had $N_\mathrm{prog}>2$. \apc{Likewise, \citet{Monachesi2019} found a median $N_{90}\simeq6.5$}. \apc{On the basis of the models we examine here}, a plausible estimate would be that the real Milky Way halo comprises dominant contributions from two to five progenitors of similar mass, and perhaps more, even if it is among the less diverse examples of a $10^{12}\,\mathrm{\msun}$ system.
 
\begin{figure}
     \centering
     \includegraphics[width=\columnwidth]{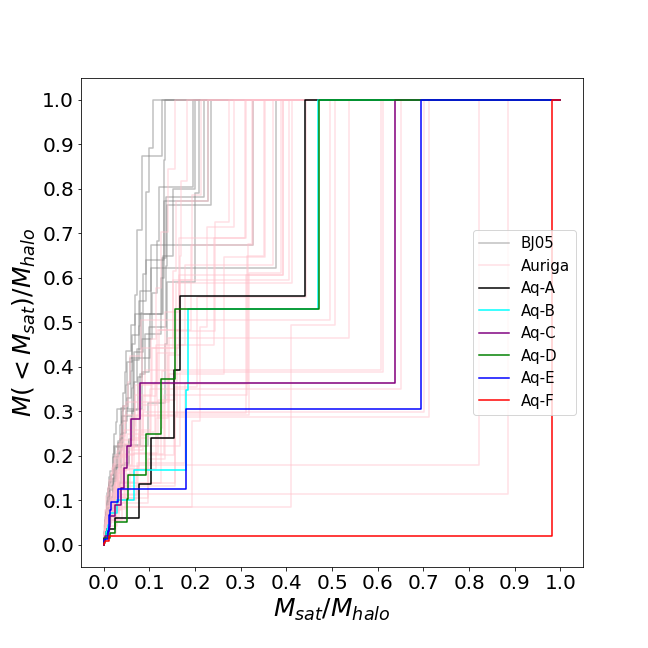}
     \caption{Cumulative mass fraction of each \replyca{accreted} stellar halo originating in progenitor satellites of stellar mass less than $M_\mathrm{sat}$ (shown on the horizontal axis normalized to the total \replyca{accreted} stellar halo mass $M_\mathrm{halo}$ in each system). Progenitor masses are sorted from smallest to largest.}
     \label{fig:cumulativemass}
 \end{figure}

Figure~\ref{fig:cumulativemass} shows the full progenitor mass distribution of each halo, scaled in a way that isolates differences in the shape of these distributions. The horizontal axis shows the mass of each progenitor, expressed as a fraction of the total accreted mass. Progenitors are sorted along this axis, with the least massive on the left and the most massive on the right. The vertical axis shows the cumulative fraction of the total accreted mass provided by each progenitor. The figure therefore shows how a given halo is built up by the contributions of individual progenitors, in rank order by mass. Halos dominated by a single very massive progenitor (for example, Aquarius halo F) show a single large step; halos built up by many small contributions (typically the case for the BJ05 halos) show many small steps. The $N_{90}$ diversity measure is equivalent to the number of steps required to pass a line at $0.9$ on the vertical axis of this figure.

Figure~\ref{fig:cumulativemass} figure helps further to explain the differences between the diversity measures shown in Figure~\ref{fig:N90comparison}. $N_\mathrm{sig}$ is sensitive to a small number of disproportionately massive progenitors, which Figure~\ref{fig:cumulativemass} shows to be typical of the Auriga and Aquarius halos. In such cases, $N_{90}$ will give relatively more weight to smaller progenitors. In this paper, we use $N_\mathrm{sig}$ to quantify the diversity of accreted halos, because it more closely reflects the way in which observational estimates proceed \apc{in the Milky Way}, starting from an approximate total mass, and weighted toward the most massive components, because those components are the most readily detected. $N_\mathrm{sig}$ is more robustly estimated from the incomplete data available in real observations; $N_{90}$ requires both an accurate estimate of the total \replyca{accreted} stellar halo mass and accurate accounting of the individual contributions from low-mass progenitors. \replyca{Together with an estimate of the dominant progenitor mass from the peak metallicity of the halo stars, stellar population gradients may help to constrain $N_{90}$ \citep{Monachesi2019}; this approach works best where $N_{90}$ is small, and is likely to be more useful for external galaxies, where population gradients can be inferred from the color of the integrated light.}
 
 \subsection{Variation of diversity with galactocentric radius}
 \label{sec:radial_diversity}
 
\begin{figure*}
     \centering
     \includegraphics[width = 1\linewidth]{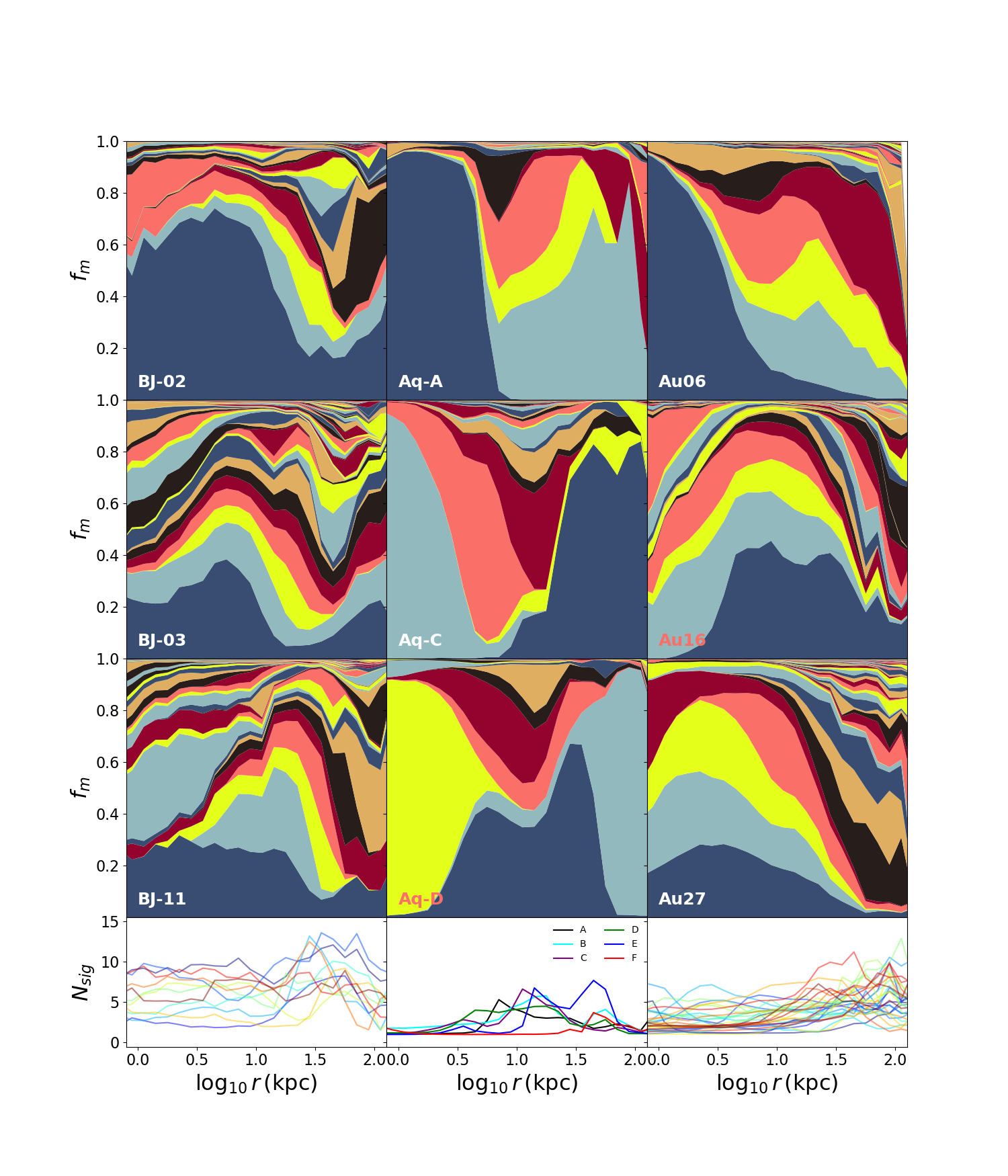}
     \caption{The strata plots and diversity profiles of each simulation over 1-100 kpc. The left panel consists of BJ05 halos, the middle panel shows Aquarius halos, and the right panel displays Auriga halos. For each simulation, the upper three panels are the strata plots; the lowest panel is the diversity profile of each halo, where the diversity is defined by \citet{Cooper2017}. The strata plots show the fraction of \replyca{accreted} stellar halo mass contributed by individual progenitors at different galactic radii. Curves for different progenitors are stacked to form a strata look. The diversity profiles quantify the strata plots and show the number of significant progenitors on different radii (i.e. the more layers a strata plot has at the certain radius, the higher the $N_{sig}$ or the diversity is).}
     \label{fig:strata-diversity}
 \end{figure*}

 Figure~\ref{fig:strata-diversity} shows the main results of this paper: the separation of the total accreted stellar halo mass at a given radius into the fractional contributions from each progenitor. We call the diagrams in the top three rows of this figure `strata plots' because of their passing resemblance to illustrations of geological strata. Contributions from a single progenitor over different radial bins (here ranging over $0$ to $\sim100\,\mathrm{kpc}$) are shaded with the same color, and they have been sorted vertically to place larger contributions relatively lower on the vertical scale of the diagram\footnote{Unlike geological stratigraphy, the absolute vertical `height' of a layer on these diagrams has no physical significance. The different contributions to the \replyca{accreted} stellar halo are well-mixed in configuration space.}. \replyca{\citet{H3} used this diagram to illustrate their interpretation of observational constraints on the mix of stellar halo progenitors in the Milky Way; it has since been used in studies of the Auriga simulations by \citet{Fattahi:2020aa} and \citet{Orkney:2023aa}}. We will discuss the Milky Way data in the next section, after exploring similarities and differences between the strata plots from our three model sets.

The bottom row of Figure~\ref{fig:strata-diversity} shows $N_\mathrm{sig}(r)$, computed in the same radial bins as the strata plots. These curves, which we call `radial diversity profiles,' summarize the distribution of progenitor contributions to the total mass at each radius. The BJ05 halos mostly have high $N_\mathrm{sig}$ at all radii, although some have notably less diversity within $\sim10\,\mathrm{kpc}$. The Aquarius halos have lower diversity overall. Except for very low diversity ($N_\mathrm{sig}\sim 1$) in their centers, there is little in common between the profiles of the different Aquarius halos. In contrast to the other models, the Aquarius diversity profiles peak in the range $10$--$80\,\mathrm{kpc}$, falling at larger radii. The diversity profiles of the Auriga halos range between these two extremes. \replyca{Although Auriga contains examples of halos with low diversity at smaller radii, similar to Aquarius, it also contains many with higher diversity in the same region}. \replyjn{It is hard to say whether this is a systematic difference between Auriga and Aquarius, or simply due to the lack of halos with high diversity in the small Aquarius sample}. 
\replyca{However}, unlike the Aquarius profiles, the diversity of most Auriga halos at $\gtrsim 100\,\mathrm{kpc}$ is similar to their peak value, or still rising.

The strata plots provide further information on the origin of these trends. A smaller number of relatively massive progenitors dominate the Aquarius and Auriga strata, in contrast to most BJ05 halos. The lower diversity of Aquarius and Auriga overall is apparent, as is their trend of increasing diversity with radius. From the strata, we can further see that the rise in diversity with radius is usually driven by the relatively concentrated radial extent of debris from the dominant central progenitors, consistent with their early accretion and/or rapid inspiral due to dynamical friction prior to disruption \citep[e.g.][]{Amorisco:2017aa}. The lower diversity in the outskirts of the Aquarius halos, which is not reflected in the majority of Auriga galaxies, is due to the contribution of a small number of comparably massive systems. This gives rise to apparent peaks in diversity at intermediate radii, where these progenitors do not dominate. This appears to be a systematic difference between the Aquarius and Auriga galaxies, with the latter tending to have significantly more contribution from less massive progenitors in their outskirts (resembling the BJ05 galaxies, although with fewer such progenitors).
 
 By design, all the BJ05 halos must grow to the same virial mass, $1.4\times10^{12}\text{M}_\odot$, without any recent major mergers. This makes it likely that they are biased toward assembly histories that involve many low-mass progenitors, which is apparent also in Figs.~\ref{fig:N90comparison} and \ref{fig:cumulativemass}. Since the individual progenitors are less massive, each makes a smaller fractional contribution, giving rise to higher diversity, in particular in the innermost regions of the halos. In these halos, the debris from many less massive progenitors can be traced over a wide range of radius (as strata of small but roughly constant thickness). This may be related to assumptions about the satellite initial conditions, progenitor orbits, or dynamical friction in BJ05.
 
The overall impression given by Figure~\ref{fig:cumulativemass} is that range of diversity among the Aquarius halos is similar to that of the Auriga halos; however, the diversity profiles of individual halos in Figure~\ref{fig:strata-diversity} show \replyca{possible systematic} differences between these two sets of simulations. 

Since Aquarius and Auriga are both cosmological $N$-body simulations of similar Milky Way–like systems, it seems unlikely that \replyca{these differences} arise only from halo-to-halo variations in progenitor mass or accretion time. Nevertheless, the star formation histories, and consequently the stellar masses, of the progenitors may be quite different in the two sets of models. For example, if a given model forms systematically more stars in progenitors that are accreted at early times, that may be reflected in the diversity profiles. Differences between particle tagging (Aquarius) and more self-consistent hydrodynamics (Auriga) may also be relevant. These factors were discussed by \citet[][]{Cooper2017}, who showed (for a single example, Aq-C) that differences in star formation histories predicted by semi-analytic and hydrodynamical models applied to the same initial conditions were likely to dominate differences in the diversity profile, while the limitations of particle tagging contributed in more subtle ways. The absence of a massive central disk potential in Aquarius may be particularly important, because such a potential could strongly affect the orbits of the most massive progenitors \replyca{(for example, progenitors may be disrupted earlier, and their orbits may be circularized in the plane of the disk)}. In that regard, \replyca{we note} that the BJ05 simulations, which include a (rigid) disk potential, do not recover behavior resembling that of the Auriga galaxies in the central regions of the halo, \replyca{although this may be due to other differences between BJ05 and Auriga.}

 
 \subsection{Comparison to the Milky Way}
 \label{sec:compare_mw}
 
 In this section, we compare the simulations above to current constraints on the composition of the Milky Way's \replyca{accreted} stellar halo. \citet{H3} present a census of likely contributions to the MW stellar halo in the range 8--45~kpc, based on the literature and their own measurements from the H3 spectroscopic survey. This radial range corresponds to Figure~19 of \citet{H3}, which directly inspired our `strata plot' presentation. \replyca{\citet{H3} attempt to correct their spectroscopic sample for various sources of incompleteness in their selection function, but \textit{not} for possible bias due to limited sky coverage; their data are restricted to a high-latitude footprint—and, within that footprint, to heights above the disk plane $|Z| > 2\,\mathrm{kpc}$. It is possible that some progenitors may make significant contributions to the volume that \citeauthor{H3} do not probe. For example, \citet{Gomez:2017aa} discuss examples from simulations in which much of the debris from massive progenitors is deposited in the plane of the disk.}
 
 \replyca{To simplify and generalize the presentation here, in particular with regard to comparisons between the different models, we do not impose these geometric limits on the simulations. This simplification should be kept in mind when assessing the differences between the simulations and the results of \citeauthor{H3}}\footnote{\replyca{\citet{Kizhuprakkat:2024ab} constructed mock catalogs for the DESI Milky Way Survey, which has a qualitatively similar footprint to H3, based on the L3 Auriga simulations. They find that the 10 most massive accreted progenitors are visible in such a footprint, regardless of the azimuthal position of the Sun in the simulated galaxy. Moreover, their relative contributions in the DESI footprint closely match their relative contributions to the halo overall.}}
 
 \citet{H3} present their interpretation in terms of mass fractions rather than absolute masses. Our definition of $N_\mathrm{sig}$ in equation~\ref{eq:1} can be rewritten in terms of mass fractions:
 \begin{equation}
     \tilde{N}_{\mathrm{sig}}(r) = \Big[\sum_i~f_{m_i}(r)\Big]^2\Big/\sum_i\Big[f_{m_i}(r)^2\Big]
     \label{eq:2}
 \end{equation}
 where $f_{m_i}(r)$ is the fraction of (observed) mass associated with the $i$th progenitor at radius $r$. This is almost equivalent to our first definition; however, in observations, the total mass at a given radius is unknown and some progenitors may be undetected.
 
  \begin{figure}
     \centering
     \includegraphics[width=\columnwidth]{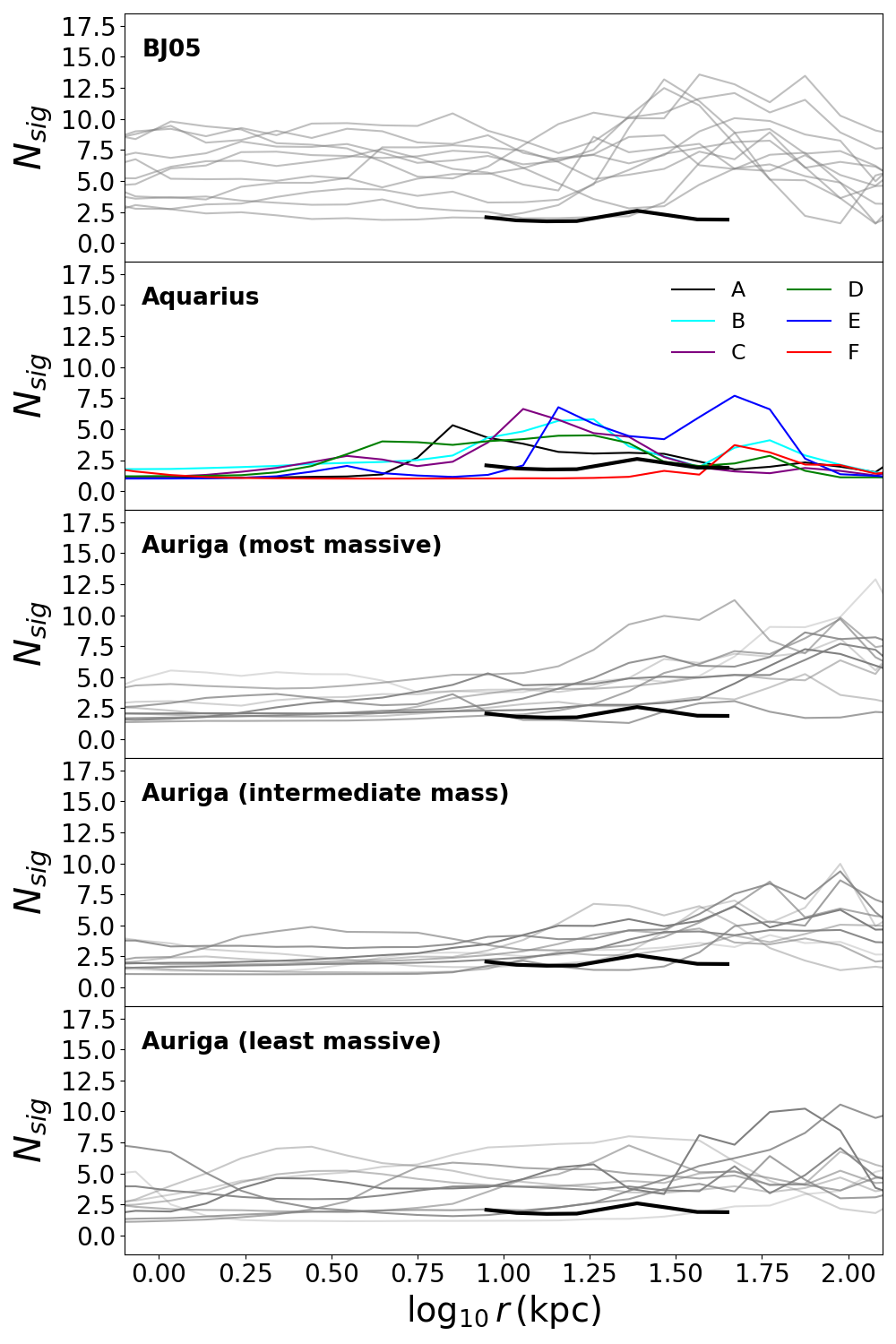}
     \caption{The diversity profile of \replyca{accreted} stellar halos. From top to bottom, panels show the BJ05 halos, the Aquarius halos, and the Auriga halos, split into three groups according to the virial mass of each galaxy. Thicker and darker gray lines indicate relatively higher \replyca{accreted} stellar halo mass in each group. The shorter black line, repeated in every panel, shows the Milky Way diversity profile we have computed based on the data in \citet{H3}.}
     \label{fig:nsig}
  \end{figure}
  
  In Figure~\ref{fig:nsig}, we compare the diversity profiles of selected examples from the models to that of the Milky Way, which we computed based on the information given in Figure 19 of \citet{H3}. We carry out this comparison at face value, without examining the assumptions underlying the interpretation of \citet{H3}. We have excluded from our comparison the fraction of mass they associate with in situ stars. Surprisingly, we find the implied diversity of the Milky Way is low compared to almost all of the models. In contrast to most simulated halos, 75\% of the Milky Way halo appears to be contributed by only two massive progenitors, \gse{} and Sagittarius, across the \replyjn{radial} range we study\footnote{\replyjn{Some potentially significant debris features that have been proposed are not currently identified in the H3 sample, notably those in the inner region of the galaxy \citep[e.g.][]{Kruijssen:2018,Horta:2020,Kruijssen:2020aa}. It is not yet clear if these are separate features and distinct from the in situ halo in the inner galaxy \citep[see, e.g.,\ section 4.3.2 of][]{Deason:2024aa}. \citet{Horta:2024aa} find analogous features in simulations are associated with progenitors that deposit the bulk of their mass within $<10$~kpc. We do not expect the absence of these features in H3 to have a strong effect on our comparison with \citet{H3}. If (for example) some fraction of the in situ \citet{H3} is really accreted debris associated with an inner-Galaxy structure, this would give only a small increase in diversity up to $\sim15$~kpc.}}.
  
\begin{figure}
     \centering
     \includegraphics[width=\columnwidth]{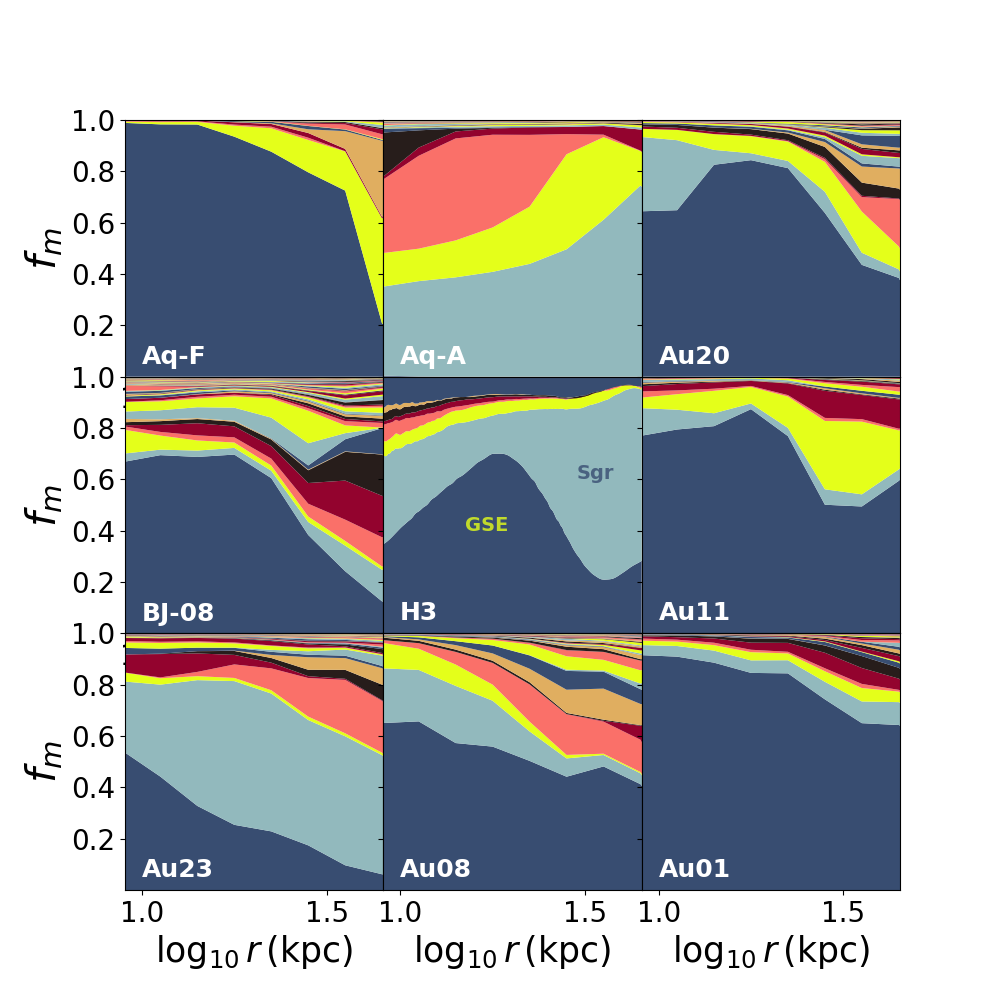}
     \caption{A subset of strata plots for halos that have low diversity and `flat' diversity profiles to $\sim40$~kpc, resembling current inferences for the Milky Way (central panel, see text). The majority of halos selected in this way experience substantial (low mass ratio) mergers close to the present day (see Figure~\ref{fig:h3comparison with time}).}
     \label{fig:h3comparison}
  \end{figure}

  In Figure~\ref{fig:h3comparison}, we show strata plots in the range $8-45$~kpc for a subset of simulated halos from the three models that we judge (by eye) to have diversity profiles most similar to that of the Milky Way (in Figure~\ref{fig:nsig}). We again compare to (a slightly modified version of) Figure~19 from \citet{H3}, shown in the central panel\footnote{Note that the radii in this figure are galactocentric, not heliocentric. In this reconstruction, the contribution of the Sagittarius stream at radii $<10$~kpc is on the opposite side of the Galactic center from the Sun.}.

These `best-matching' halos are among the least typical in each model set. As shown in Figure~\ref{fig:cumulativemass}, the most massive progenitor of Aq-F contributes around 98\% of the total \replyca{accreted} stellar halo mass. This galaxy has undergone a recent major merger and hence is unlikely to have an undisturbed disk resembling the Milky Way. In projection, its halo is dominated by a prominent system of `shells' \citep[see][]{Cooper:2011ab}.  Similarly, \citet{Fattahi2019} exclude Au11 and Au20 from their study on the basis that they are still undergoing merger events at $z = 0$. \citet{Monachesi2019} exclude Au11 and Au01, which they find to have a nearly equal-mass companion nearby. These four halos could therefore be said to be among the least `Milky Way—like' of the Auriga set, due to their recent or ongoing merging. In contrast to these examples, the Milky Way seems to have atypically low diversity, despite (apparently) not having undergone a recent major merger. 

  \begin{figure}
      \centering
      \includegraphics[width=\columnwidth]{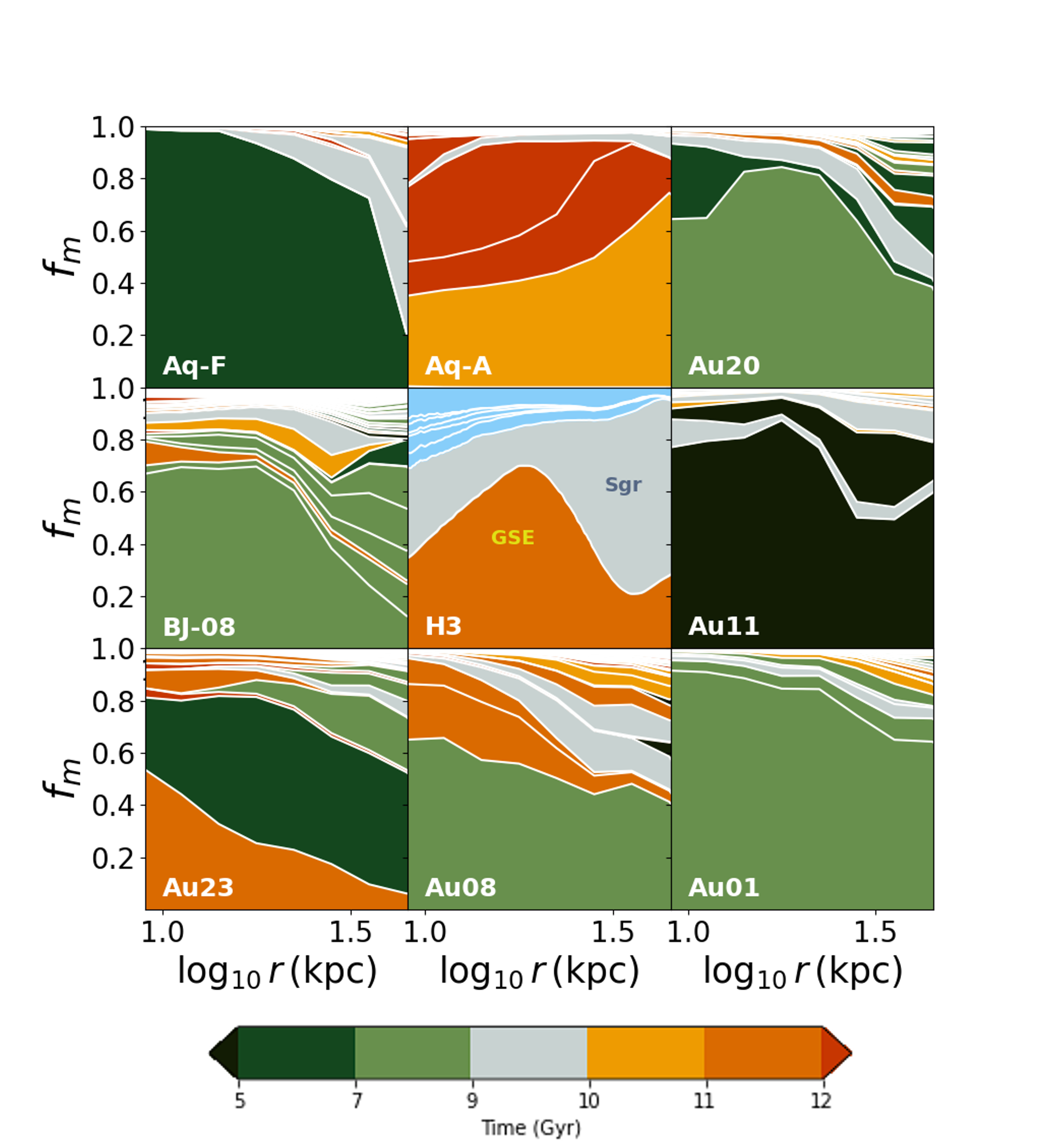}
      \caption{As Figure~\ref{fig:h3comparison}, but here with strata colored according to the infall \replyjn{(lookback)} time of each progenitor. The estimated infall times of \gse{} and Sgr are taken from \citet{GSE_time_reference} and \citet{Sgr_time_reference}, respectively. We color the other Milky Way progenitors blue to indicate that their infall times are not yet well constrained.}
      \label{fig:h3comparison with time}
  \end{figure}
  
The discrepancy between the Milky Way and the simulations is even clearer in Figure~\ref{fig:h3comparison with time}, where we color code the strata according to the time of infall of the progenitor (an upper limit to the lookback time at which the stars were stripped into the halo). The current consensus is that \gse{} originated in an `early' merger \citep[$\gtrsim 11\,\mathrm{Gyr}$ ago, e.g.,][]{GSE_time_reference}. Although such early mergers do occur in the models (e.g.,\ in Au23), they are rarely dominant contributors to the present-day stellar halo. When a single progenitor dominates the present-day halo, that progenitor typically arrives relatively late 
($\lesssim9$~Gyr ago), as in Au20, Au11, Au01, Aq-F, and BJ08. 

\citet{Fattahi2019} searched the Auriga halos for features that resemble the observed spatial, chemical, and kinematic properties of \gse{} \replyca{\citep[see also][]{Orkney:2023aa}}. 
None of the Auriga halos in Figure~\ref{fig:h3comparison} are \gse{} analogs according to their \replyca{criteria}. Conversely, none of the \gse{}-analog Auriga halos identified by \citet{Fattahi2019} have diversity profiles resembling that of the Milky Way. Superficially `\gse-like' (early-accreted, centrally concentrated, high mass fraction) features in the strata plots of halos with somewhat Milky Way–like diversity profiles (for example, Au08) are not \gse-like according to the criteria of \citet{Fattahi2019}. Furthermore, we do not see any notable similarity among the strata plots or diversity profiles of the \citet{Fattahi2019} \gse{} analogs. 
 
Essentially the same considerations apply in the case of Sgr; the only difference is a somewhat later infall time and a more circular orbit, leading to more deposition in the outer halo. The infall time of Sgr is thought to be $\sim 9\,\mathrm{Gyr}$ ago \citep[e.g.][]{Sgr_time_reference}, and the progenitor has not yet been completely disrupted. In a halo dominated by a single massive \gse-like progenitor, the late arrival of a comparably massive Sgr-like progenitor could produce a diversity distribution that resembles the picture sketched for the Milky Way. Might the Milky Way have been more consistent with the models before Sagittarius arrived? At face value, the answer is no -- the models we have examined suggest that a late-type galaxy with an \replyca{accreted} stellar halo dominated by a single \gse{} analog would be even more unusual. The state after the accretion of a late-arriving Sgr analog is arguably closer to typical. 

Therefore, taking the models at face value, the Milky Way appears unusual. We do not find any halos out of the 47 Milky Way analogs from these three different models that resemble the combination of apparently low diversity and early assembly inferred for the Milky Way. Among those simulated galaxies that do not experience a major merger, few (essentially none in our sample) have two such dominant progenitors as \gse{} and Sgr arriving so early. Moreover, the latest-arriving progenitor is usually the most massive. The only clear exception is Aq-A, an atypically early-forming dark matter halo, in which a relatively early merger (the most massive) dominates the \replyca{accreted} stellar halo alongside two even earlier mergers of comparable mass. Removing one of the two more ancient Aq-A progenitors (or interpreting observations of their debris as a single event) would produce a somewhat Milky Way–like result\footnote{These mergers in Aq-A also involve a substantial fraction of stars on a wide, somewhat Sgr-like orbit; see \citet{Cooper2010}.}. \replyjn{This apparent inconsistency with simulations also motivates consideration of more recent accretion times for the dominant Milky Way halo structures. For example, \citet{Donlon:2019aa, Donlon:2020aa, Donlon:2022aa}, \citet{Donlon2023rf} and \citet{Donlon:2024aa} examine an alternative scenario in which the \gse{} progenitor (or some fraction of the debris associated with \gse{}) merged with the Milky Way in the last 2-3~Gyr. Although this is even more recent than most of the dominant progenitors in the models we consider, a \gse{}-like accretion in the last $\sim5$~Gyr would be consistent with the overall trend in the simulations \citep[see also][]{Folsom:2024aa}}.


The statistics of this comparison are quite poor; the Milky Way is known to be an atypical galaxy for its mass in several respects (for example, the presence of the LMC \replyjn{\citep[e.g.][]{Boylan-Kolchin:2010aa,Liu:2011,Shao:2018,Christoph:2021}}, and its \replyca{apparently} relatively efficient in situ star formation \citep{Licquia:2015ww,Licquia:2016aa,Boardman:2020aa}). If being an outlier in these respects correlates with \replyca{accreted} stellar halo diversity,\footnote{A reasonable speculation, because it almost certainly reflects aspects of the halo growth history.} it is not surprising that we do not find a good match in a small sample of models based on this single point of comparison. This is compounded by the fact that the models span a wide range of virial mass, and by uncertainty on the true virial mass of the Milky Way. Our sample may include only a handful of galaxies (or none) that are appropriate mass analogs of the Milky Way.

Alternatively, the low diversity inferred for the Milky Way may be an underestimate. Other debris systems that have not been accounted for may contribute at a significant level. Perhaps the most likely reason for this underestimate would be that multiple systems have been \replyjn{identified} with a single hypothetical progenitor in the current interpretation \replyjn{of the Milky Way data}. \replyjn{This has been proposed for both \gse{} \citep[e.g.][]{Donlon:2022aa,Donlon:2024aa,Sharpe:2024aa,Carrillo:2024aa} and Sgr \citep[e.g.][]{Newby2015,Davies2024}. \citet{Folsom:2024aa} posit that $\sim1/3$ of \gse{}-like debris systems in the TNG50 simulations had multiple progenitors.}

\begin{figure}
      \centering
      \includegraphics[width=\columnwidth]{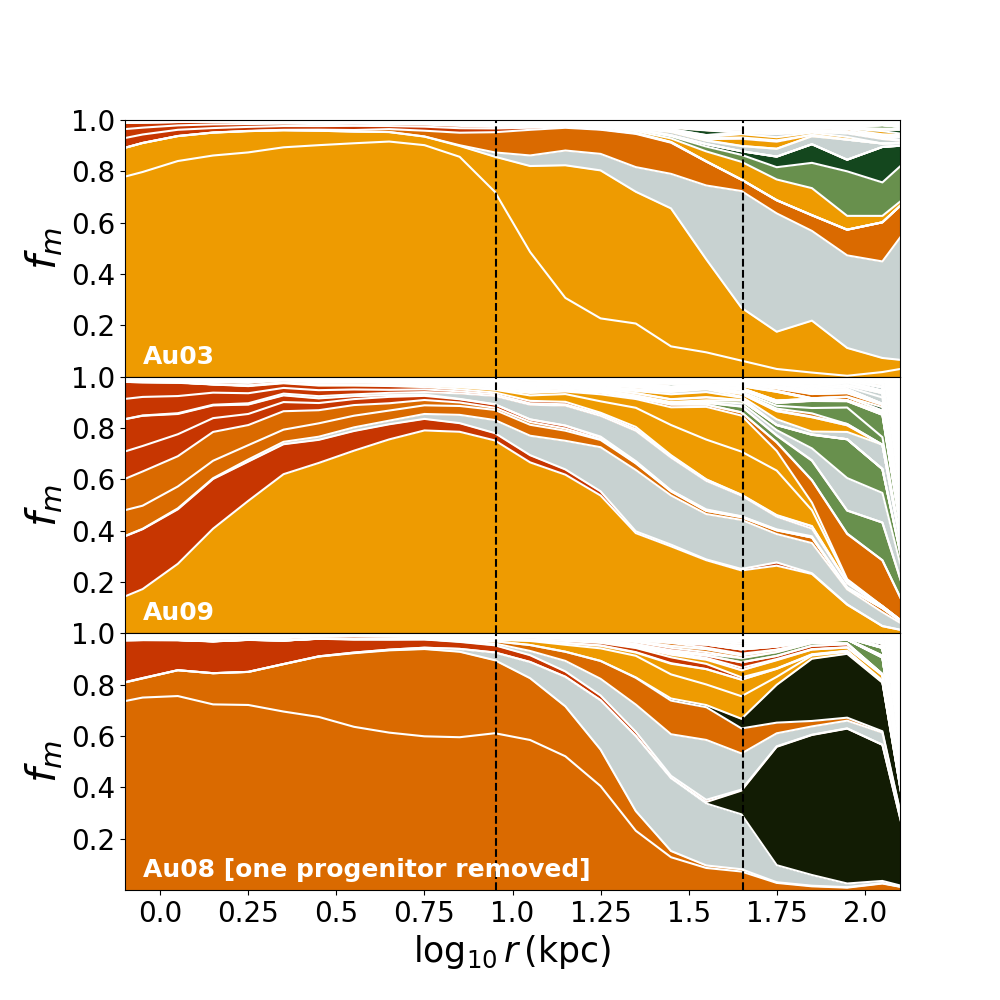}
      \caption{As Figure~\ref{fig:h3comparison with time}, but here showing three halos chosen by eye to resemble the Milky Way's apparent combination of low diversity and early accretion times for the major progenitors. Dashed lines indicate the radial range of  Figs.~\ref{fig:h3comparison} and \ref{fig:h3comparison with time}. The lowest panel shows Au08 with its most massive progenitor excluded; compare with the Au08 panels in Figs.~\ref{fig:h3comparison} and \ref{fig:h3comparison with time}. This excluded progenitor has a recent infall time; without it, Au08 much more closely resembles the Milky Way.}
      \label{fig:MWcolor}
  \end{figure}

To illustrate this point, the upper two panels of Figure~\ref{fig:MWcolor} show strata plots for Au03 and Au09, two halos with higher diversity than the Milky Way but a similar \replyca{accreted} stellar halo assembly history -- that is,  one or more progenitors of dominant, \gse-like centrally concentrated debris with infall time 10-11~Gyr ago and one or more progenitors of less concentrated Sgr-like debris with infall time 9-10~Gyr ago. \replyca{Au09 also has an \replyca{accreted} stellar halo mass and a metallicity gradient similar to those of the Milky Way \citep{Monachesi2019}}. In a comparable range of radius, we find these halos have strata much more closely resembling those of \citet{H3} in Figure~\ref{fig:h3comparison with time}. An interpretation of the Milky Way data in which `\gse{}' or `Sgr' debris was attributed to more than one progenitor with \replyca{similar masses and infall times} would be comparable to these models. 

As above, we can consider the effect of stochastic accretion on this comparison; the lowest panel of Figure~\ref{fig:MWcolor} shows the Au08 strata after removing the contribution of the dominant recent accretion event ($\lesssim 9$~Gyr ago) removed. The remaining strata are the closest match to the Milky Way interpretation of \citet{H3} that we can identify in our models—the only difference being that both the `\gse{}' analog and the `Sgr' analog each comprise two systems with similar radial distribution.

Figure~\ref{fig:MWcolor} is plotted over a larger range of radius; although \citet{H3} probe distances up to $\sim40$~kpc with sufficient density to attempt estimates of the mass fractions of different halo structures, this is still a relatively small scale in terms of the \replyca{accreted} stellar halo overall. 
Although the strata of the three examples are similar in the region probed by H3, Au09 has many more strata (higher diversity) than Au03 and Au08 at radii $<8$~kpc and $>40$~kpc. 
The outer region of Au08 is dominated by debris associated with a very recently accreted satellite (the black `hills' around $\sim100$~kpc).
Most simulated diversity profiles rise at larger radii in Auriga (although not in Aquarius). Evidence for or against the increasing diversity of the Milky Way halo at $60$-$100$~kpc would therefore provide another useful point of comparison with models.

\section{Conclusions}
\label{sec:conclusions}

We have explored the diversity of the present-day \replyca{accreted} stellar halos of Milky Way mass galaxies by quantifying the fraction of mass contributed by different progenitors as a function of galactocentric radius, in three different sets of cosmological simulations: BJ05, Aquarius, and Auriga. These models cover a range around the likely virial mass of the Milky Way (see Figure~\ref{fig:1}) and use very different methods for populating progenitor dark matter halos with stars. Each model has been used extensively in the literature to predict and interpret observations of the real Milky Way. We find systematic differences in the composition of \replyca{accreted} stellar halos predicted by these models, notwithstanding the galaxy-to-galaxy differences within any given model owing to the stochastic nature of progenitor masses, accretion times, and orbits. Specifically, we find:

\begin{enumerate}

\item The BJ05 halos have a much larger number of significant progenitors than the other two simulations, both globally (Figs.~\ref{fig:N90comparison} and ~\ref{fig:cumulativemass}) and as a function of radius (Figure~\ref{fig:nsig}). This higher diversity, reflecting assembly from a relatively larger number of lower-mass progenitors, is most likely due to assumptions made in the generation of the BJ05 models. For example, at a fixed progenitor virial mass, the BJ05 models predict stellar masses for the most massive progenitors up to an order of magnitude lower than in Auriga (Figure~\ref{fig:SMHMrelation}). 
    
\item The Aquarius halos have lower diversity within $\sim10$~kpc, reflecting dominant contributions made by (typically) only one or two progenitors. 

\item The Auriga simulations are broadly similar to the Aquarius models in the region of the \replyca{accreted} stellar halo probed by current observations, but they predict increasing diversity at a larger radius, whereas the Aquarius simulations predict a peak in diversity around $\sim50$~kpc. These differences may again be a consequence of systematic differences in the star formation histories of progenitors (the Aquarius progenitors form stars less efficiently than those in Auriga, producing less massive \replyca{accreted} stellar halos overall), combined with differences in the tidal disruption process between the hydrodynamical and particle-tagging method \citep[for a detailed discussion, see][]{Cooper2017}.

 \item There is no obvious difference in diversity between those Auriga halos that were identified by \citet{Fattahi2019} as having \gse-like progenitors and those that are not, suggesting that the chemodynamical identification of such a progenitor in the Milky Way may not translate in a simple way to expectations for the overall composition of the \replyca{accreted} stellar halo.
\end{enumerate}

We have made a preliminary, face-value comparison between the halo diversity predicted by these simulations and that implied by the Milky Way stellar halo census of \citet{H3}. We find:

\begin{enumerate}
\item Essentially none of the simulations reproduce the combination of very low diversity and early assembly implied by the Milky Way data.
\item The BJ05 simulations are generally the most discrepant because they predict the highest diversity, both locally and overall.

\item The majority of systems with diversity 
as low as that implied for the Milky Way, in the currently observed region of the halo, are examples of relatively recent accretion of a massive progenitor. \replyjn{This motivates further consideration of the accretion times of \gse{} and Sgr}.


\item There are several simulated galaxies that have broadly similar progenitor mass deposition profiles (`strata') and corresponding progenitor infall times to those inferred for Milky Way, \replyca{in particular in Auriga}. However, in these cases, the number of dominant progenitors is higher than the $\sim2$ currently suggested for the Milky Way. On this basis, the simulations predict that other significant progenitor debris systems, closely resembling the chemodynamical structure of \gse{}, await identification in the Milky Way.
\end{enumerate}
\replyca{An important caveat for the work presented here is that the \citet{H3} result refers to a specific, nontrivial survey geometry, whereas (for simplicity and generality) our analysis of the simulations refers to a complete spherical volume. The above conclusions could be changed if the \citet{H3} result is not representative of the entirety of the Milky Way's accreted stellar halo in the same radial range—for example, if accreted debris on circular orbits in the plane of the Milky Way disk makes a significant contribution \citep[e.g.][]{Gomez:2017aa}.} \replyjn{The apparent discrepancy with the simulations would also be reduced if the total stellar mass of \gse{} or Sgr was found to be an overestimate \citep[e.g.][]{Lane:2023aa}.} 

\replyca{As we describe in the introduction, most work on the diversity of accreted stellar halos predicts that $\sim1$--$10$ progenitors dominate the total accreted mass. This corresponds well} to currently available evidence for the diversity of the real Milky Way halo \replyca{\citep[and the halos of other Milky Way–like galaxies, e.g.,][]{Monachesi:2016aa, Gozman:2023aa}}. \replyca{However, the composition of the Milky Way's stellar halo may eventually be constrained in much greater detail than that of any other galaxy. In that context, the difference between 1 and 10 significant progenitors is still substantial. It is therefore worthwhile to consider the implications of the apparently low diversity of the Milky Way's stellar halo suggested by current data -- in particular, whether that finding is indicative of a particular subset of possible assembly histories.} 

\replyca{A broad range of simulations of simulated systems with comparable virial mass, using different techniques, can provide useful context and intuition when interpreting reconstructions of the Milky Way's history. In that regard, we believe it is notable that the simulations we have examined here show many cases of galaxies in the virial mass range of the Milky Way} with a larger number of significant stellar halo progenitors, by $N_\mathrm{sig}$ or other measures \citep[e.g.][]{Monachesi2019}. Moreover, halos with high diversity overall may appear to be dominated by only one or two progenitors if observed over a restricted range of radius (and vice versa). As noted by \citet{Deason2016}, preselecting galaxy assembly histories, for example to exclude major mergers, biases any conclusions regarding what kind of stellar halo assembly scenarios (and hence diversities) are typical for halos of a particular virial mass.

\citet{Deason2016} conclude that the Milky Way is atypical in hosting several particularly massive surviving satellites \citep[in particular, the LMC and SMC; see also, e.g.,][]{Cautun:2014aa} \replyca{while also having} an apparently `quiescent' accretion history, on the evidence of its stable disk and low-mass, metal-poor stellar halo. They note that this situation may be transitory; the disruption of the LMC and the SMC will, in this and other respects `restore the Milky Way to the mean' \citep{Cautun:2019aa}. A stellar halo dominated by two recently accreted progenitors with somewhat correlated orbits would resemble the example of Au08 presented above, and hence would be more typical of the low-diversity galaxies we find in the simulations.  

Our results suggest that predictions of diversity from simulations are strongly model-dependent. Clearly, it would be valuable, and relatively straightforward, to make similar predictions with the many other Milky Way–analog zoom simulations now available. Our work suggests that such predictions can be usefully compared to ongoing and future Milky Way surveys, including DESI MWS \citep{desimws}, 4MOST \citep{4most}, WEAVE \citep{Weave}, SDSS-V \citep{SDSSV}, and PFS \citep{PFS2022}. These surveys will further improve the constraints on halo pioneered by \citet{H3}, in particular by extending the radial range studied and the number of stars \replyca{observed}. 

\begin{acknowledgments}

\replyjn{We thank the anonymous reviewer for a thorough and constructive report, which greatly improved the clarity of the paper and brought important related work to our attention}. PSY and APC are supported by a Yushan Fellowship awarded by the Taiwanese Ministry of Education (MOE), MOE-113-YSFMS-0002-001-P2. APC acknowledges support from Taiwan's National Science and Technology Council (NSTC) under grants 109-2112-M-007-011-MY3, 112-2112-M-007-017 and 113-2112-M-007-009. AM gratefully acknowledges support by FONDECYT Regular grant 1212046 and by the ANID BASAL project FB210003, as well as funding from the Max Planck Society through a `PartnerGroup' grant and the HORIZON-MSCA-2021-SE-01 Research and Innovation Programme under the Marie Sklodowska-Curie grant agreement number 101086388. RG is supported by an STFC Ernest Rutherford Fellowship (ST/W003643/1). This work used high-performance computing facilities operated by the Center for Informatics and Computation in Astronomy (CICA) at National Tsing Hua University (NTHU). This equipment was funded by MOE, NSTC, and NTHU. We have used  the Auriga Project data release \citep[][]{Grand:2024aa} available at \href{https://wwwmpa.mpa-garching.mpg.de/auriga/data}{https://wwwmpa.mpa-garching.mpg.de/auriga/data}.

\end{acknowledgments}

%

\vspace{5mm}





\section*{Data Availability}

We provide the \citet{Cooper2010} star particle data for Aquarius at \url{https://doi.org/10.5281/zenodo.13888986}: this is the first public release of those data. 
A description of \replyjn{these} data and related scripts are available at \url{https://github.com/nthu-ga/aquarius-halos}, as described in the appendix.  Our progenitor IDs for the stellar particles in the Auriga data release are provided at 
\url{https://doi.org/10.5281/zenodo.13943963}.

\appendix

\section{Stellar Halo Data} \label{Appendix1}

We make the Aquarius simulation data of \citet{Cooper2010} publicly available in HDF5 format at 
\url{https://doi.org/10.5281/zenodo.13888986}. The data model for these files is described in Table~\ref{tab:datamodel} and further documented at \url{https://github.com/nthu-ga/aquarius-halos}. The dataset names and the file header follow the conventions of the Gadget file format and should be compatible with most routines designed to read that format. At the above URL, we also provide scripts to convert the \citet{B&J2005} simulation outputs to \replyjn{the same} format \replyjn{we use for Aquarius}, along with progenitor ID labels for accreted star particles in the L3 Auriga halos. We plan to update these data products with extra properties (such as progenitor infall time, metallicity, and age) and to provide other related data products in association with future publications. 

The rows of each dataset in these files correspond to `star particles' in the simulations. Only accreted particles are included. The Cartesian coordinate frame is that of the center of the galactic potential. The field \texttt{ProgenitorID} identifies the last branch of the subhalo merger tree to which a star particle was bound before it became bound to the dark matter halo of the central galaxy. These labels should be treated as arbitrary integers. Particles currently bound to satellite subhalos can be identified by \texttt{SubhaloNr $>0$}. For further details, see the documentation URL above.

\begin{deluxetable*}{lll}
    \label{tab:datamodel}
    \tabletypesize{\scriptsize}
    \tablewidth{0pt}
    \tablecaption{Data model of the HDF5 files provided for the Aquarius simulations. For Auriga, we provide only the ProgenitorID column, as the other columns are provided in the Auriga data release.}
    \tablehead{
    \colhead{Data Set} & \colhead{Unit} & \colhead{Description}
    }
    \startdata
    \texttt{Coordinates}     & $\mathrm{h^{-1}\,Mpc}$ & Galactocentric Cartesian coordinates \\ 
    \texttt{Mass}            & $\mathrm{h^{-1}\,M_{\odot}}$ & Stellar mass \\   
    \texttt{ParticleIDs}     & - & Unique integer label for each particle \\    
    \texttt{ProgenitorID}    & - & Unique integer label for each progenitor \\      
    \texttt{SubhaloNr}       & - & $0$: main halo, $>0$: satellite, $-1$: unbound \\
    \texttt{Velocities}     & $\mathrm{km\,s^{-1}}$ & Galactocentric Cartesian velocities \\
    \enddata
\end{deluxetable*}

\bibliography{example}{}
\bibliographystyle{aasjournal}



\end{document}